\documentclass[twocolumn,showpacs,preprintnumbers,amsmath,amssymb,prl]{revtex4}
\usepackage{graphicx,longtable}
\usepackage{epsf,dsfont,hyperref,url,float}

\begin{document}

\title{Experimental Validation of Quantum Game Theory}

\author{Matthias Hanauske}
\author{Steffen Bernius}
\author{Wolfgang K\"onig}
\affiliation{
Johann Wolfgang Goethe-University\\
Institute of Information Systems\\
Mertonstr. 17, 60054 Frankfurt$/$Main
}
\author{Berndt Dugall}
\affiliation{
Johann Christian Senckenberg University-Library\\
Bockenheimer Landstr. 134-138, 60325 Frankfurt$/$Main
}
\date{\today}

\begin{abstract} 
This article uses data from two experimental studies of two-person Prisoner's Dilemma games \cite{blonski_2007,dal_2006} and compares the data with the theoretic predictions calculated with the use of a quantum game theoretical method. The experimental findings of the cooperation percentage $C_p$ indicate a strong connectivity with the properties of a novel function ($\cal{N}$), which depends on the payoff parameters of the game and the entanglement parameter $\gamma$. A classification scheme depending on four quantum cooperation indicators is developed to describe cooperation in real two-person games. The quantum indicators lead to results, that are at least as good as the cooperation predictions derived from classical game theory.   
\end{abstract}

\maketitle

Quantum game theory has its origin in elementary particle physics and quantum information theory. In 1999 the first formulations of quantum game theory where presented by D. A. Meyer \cite{meyer-1999-82} and J. Eisert et al. \cite{eisert-1999-83}. Unknowing Meyers' results on the ``Penny Flip'' game, Eisert and colleagues focused on the prisoner's dilemma game. Within their quantum representation they where able to demonstrate, that prisoners could escape from the dilemma, if the entanglement of the wave functions lies above a certain value. In 2001 J. Du et al. \cite{du-2002-88} realized the first simulation of a quantum game on their nuclear magnetic resonance quantum computer. The application of quantum game theory on an existing social system, namely the open access publication network of scientists, was presented in M. Hanauske et al. \cite{hanauske_2007}. The authors showed, that quantum game theory could give a possible explanation of the differing publishing methods of scientific communities. 
A validation of quantum game theoretical concepts by using experimental data of real two-person games was addressed in K.-Y. Chen and T. Hogg \cite{chen-2006} (see also \cite{patel-2007}). In contrast to the experimental data used in the present article the authors of \cite{chen-2006} used an experimental design, which includes a quantum version of the game. Our understanding of an inclusion of quantum strategies in the players' decisions is different, insofar as we interpret the whole process of a real game as a quantum game.  

In the present paper, based on the Eisert's two-player quantum protocol \cite{eisert-1999-83} and the concept of quantum Nash equilibria, four quantum cooperation indicators are defined. By using these indicators to predict the cooperation rates of real two-person games it will be shown, that the quantum indicators lead to results, that are at least as good as the cooperation predictions derived from classical game theory.

\section{Mathematics of QGT}
The normal-form representation of a two-player game $\Gamma$, where each player (Player 1 $\hat{=}$ A, Player 2 $\hat{=}$ B) can choose between two strategies (${\cal S}^{A}=\{s^{A}_1,s^{A}_2\}$, ${\cal S}^{B}=\{s^{B}_1,s^{B}_2\}$) is the classical grounding of the two-player quantum game focused on in this article. In our case the two strategies represent the players' choice between cooperating (C) or defecting (D) in a prisoner's dilemma game. 
The whole strategy space ${\cal S}$ is composed with use of a Cartesian product of the individual strategies of the two players: 
\begin{equation}
{\cal S} =  {\cal S}^A \times {\cal S}^B = \left\{ \hbox{(C,C)}, \hbox{(C,D)}, \hbox{(D,C)}, \hbox{(D,D)} \right\} \quad .
\end{equation}
The payoff structure of a prisoner's dilemma game can be described by the following matrix:
\begin{table}[H]
\centerline{
\begin{tabular}{|r|c c|}
        \hline
        A$\setminus$B& C & D
        \\ \hline
        C & (c,c) & (a,b)
        \\
        D & (b,a) & (d,d)\\\hline
\end{tabular}}
\caption[caption]{General prisoner's dilemma payoff matrix.}
\label{tab:PayOff_general}
\end{table}
The parameters a, b, c and d should satisfy the following inequations \cite{blonski_2007}
\begin{equation}
b > c > d > a \,\, , \quad 2\,c > a+b \quad .
\end{equation}
 
In quantum game theory the measurable classical strategies (C and D) correspond to the orthonormal unit basis vectors $\left| C \right>$ and $\left| D \right>$ of the two dimensional complex space $\mathds{C}^2$, the so called Hilbert space ${\cal{H}}_i$ of the player $i$ ($i=A,B$). A quantum strategy of a player i is represented as a general unit vector $\left| \psi \right>_i$ in his strategic Hilbert space ${\cal{H}}_i$. The whole quantum strategy space $\cal{H}$ is constructed with the use of the direct tensor product of the individual Hilbert spaces: ${\cal{H}}:={\cal{H}}_A \otimes {\cal{H}}_B$. The main difference between classical and quantum game theory is, that in the Hilbert space ${\cal{H}}$ correlations between the players' individual quantum strategies are allowed, if the two quantum strategies $\left| \psi \right>_A$ and $\left| \psi \right>_B$ are entangled. The overall state of the system we are looking at is described as a two-player quantum state $\left| \Psi \right> \in {\cal{H}}$. We define the four basis vectors of the Hilbert space ${\cal{H}}$ as the classical game outcomes ($\left| CC \right>:=(1,0,0,0)$, $\left| CD \right>:=(0,-1,0,0)$, $\left| DC \right>:=(0,0,-1,0)$ and $\left| DD \right>:=(0,0,0,1)$). 

The setup of the quantum game begins with the choice of the initial state $\left| \Psi_0 \right>$. We assume that both players are in the state $\left| C \right>$. The initial state of the two players is then given by 

\begin{equation}
\left| \Psi_0 \right> \,=\, \widehat{\mathcal{J}} \left| CC \right> \,=\,
\left( \begin{array}{c}
 \cos\left( \frac{\gamma}{2} \right) \\ \\ 0 \\ \\ 0 \\ \\ i\sin\left( \frac{\gamma}{2}\right) \\
 \end{array} \right)  \quad ,
\end{equation}
where the unitary operator $\hat{\cal{J}}$ is responsible for the possible entanglement of the two-player system. The players' quantum decision (quantum strategy) is formulated with the use of a two parameter set of unitary $2\times2$ matrices: 
\begin{eqnarray}
&\widehat{\mathcal{U}}(\theta,\varphi) :=
\left(
\begin{array}[c]{cc}
e^{i\,\varphi} \, \hbox{cos}(\frac{\theta}{2})&\hbox{sin}(\frac{\theta}{2})\\
-\hbox{sin}(\frac{\theta}{2})&e^{-i\,\varphi} \, \hbox{cos}(\frac{\theta}{2})
\end{array}
\right)&\\
&
\forall \quad \theta \in{} [0,\pi] \,\, \wedge \,\, \varphi \in{} [0,\frac{\pi}{2}] &\quad .\nonumber
\end{eqnarray}
By arranging the parameters $\theta$ and $\varphi$ a player is choosing his quantum strategy. The classical strategy C for example is selected by appointing $\theta=0$ and $\varphi=0$ :
\begin{equation}
\widehat{\mathcal{C}}:=
\hat{{\cal{U}}}(0,0) =
\left(
\begin{array}[c]{cc}
1&0\\
0&1
\end{array}
\right)\quad,
\end{equation}
whereas the strategy D is selected by choosing $\theta=\pi$ and $\varphi=0$ :
\begin{equation}
\widehat{\mathcal{D}}:=
\hat{{\cal{U}}}(\pi,0) =
\left(
\begin{array}[c]{cc}
0&1\\
-1&0
\end{array}
\right)\quad.
\end{equation}
In addition, the quantum strategy $\widehat{\mathcal{Q}}$ is given by
\begin{equation}
\widehat{\mathcal{Q}}:=
\hat{{\cal{U}}}(0,\pi/2) =
\left(
\begin{array}[c]{cc}
i&0\\
0&-i
\end{array}
\right)\quad.
\end{equation}

After the two players have chosen their individual quantum strategies ($\hat{{\cal{U}}}_A:=\hat{{\cal{U}}}(\theta_A,\varphi_A)$ and $\hat{{\cal{U}}}_B:=\hat{{\cal{U}}}(\theta_B,\varphi_B)$) the disentangling operator $\widehat{\mathcal{J}}^\dagger$ is acting to prepare the measurement of the players' state. The entangling and disentangling operator ($\widehat{\mathcal{J}}, \widehat{\mathcal{J}}^\dagger$; with $\hat{\cal{J}}\equiv\hat{\cal{J}}^\dagger$) is depending on one additional single parameter $\gamma$ which measures the strength of the entanglement of the system:
\begin{equation}
\widehat{\mathcal{J}} := e^{i \, \frac{\gamma}{2} (\widehat{\mathcal{D}} \otimes \, \widehat{\mathcal{D}})} \,\, , \quad \gamma \in{} [0,\frac{\pi}{2}] \quad .
\end{equation}
The entangling operator $\widehat{\mathcal{J}}$ in the used representation has the following explicit structure:  
\begin{equation}
\widehat{\mathcal{J}} := \left( \begin{array}{cccc}
 \cos\left( \frac{\gamma}{2} \right) & 0 & 0 & i\sin\left( \frac{\gamma}{2} \right) \\
  & & & \\
 0 & \cos\left( \frac{\gamma}{2}\right) & -i\sin\left( \frac{\gamma}{2}\right) & 0 \\
  & & & \\
 0 & -i\sin\left( \frac{\gamma}{2}\right) & \cos\left( \frac{\gamma}{2}\right) & 0 \\
  & & & \\
 i\sin\left( \frac{\gamma}{2}\right) & 0 & 0 & \cos\left( \frac{\gamma}{2}\right) \\
 \end{array} \right)
\end{equation}

Finally the state prior to detection can therefore be formulated as follows:
\begin{equation}
\left| \Psi_f \right> = \hat{\cal{J}}^\dagger \left( \hat{\cal{U}}_A \otimes \hat{\cal{U}}_B \right) \hat{\cal{J}}\, \left| CC \right> \quad .
\end{equation}
The expected payoff within a quantum version of a general two-player game, depends on the payoff matrix (see Table \ref{tab:PayOff_general}) and on the joint probability to observe the four possible outcomes of the game:  
\begin{eqnarray}
&\$_A&= c\,P_{\mbox{\small  CC}} + a\,P_{\mbox{\small CD}} + b\,P_{\mbox{\small DC}} + d\,P_{\mbox{\small DD}} \\
&\$_B&= c\,P_{\mbox{\small CC}} + b\,P_{\mbox{\small CD}} + a\,P_{\mbox{\small DC}} + d\,P_{\mbox{\small DD}} \nonumber\\
&\mbox{with:}& P_{\sigma \sigma^{,}}=\left| \, \left< \sigma\sigma^{,} | \Psi_f \right> \, \right|^2 \,\, , \quad \sigma,\sigma^{,}=\left\{ C,D \right\} \quad . \nonumber
\end{eqnarray}

\section{Quantum Cooperation Indicators}

Dominant quantum strategies and quantum Nash equilibria are formulated as follows:\\  

($\theta_A^\star,\varphi_A^\star;\theta_B^\star,\varphi_B^\star$) is a dominant quantum strategy, if
\begin{eqnarray}
\$_A(\widehat{\mathcal{U}}_A^\star,\widehat{\mathcal{U}}_B) &\geq& 
\$_A(\widehat{\mathcal{U}}_A,\widehat{\mathcal{U}}_B)\qquad   
\forall \quad
\widehat{\mathcal{U}}_A\,\, \wedge \,\,\widehat{\mathcal{U}}_B \\
\$_B(\widehat{\mathcal{U}}_A,\widehat{\mathcal{U}}_B^\star) &\geq& 
\$_B(\widehat{\mathcal{U}}_A,\widehat{\mathcal{U}}_B) \qquad   
\forall \quad
\widehat{\mathcal{U}}_A\,\, \wedge \,\,\widehat{\mathcal{U}}_B \quad .
\nonumber
\end{eqnarray}

($\theta_A^\star,\varphi_A^\star;\theta_B^\star,\varphi_B^\star$) is a quantum Nash equilibrium, if
\begin{eqnarray}
\$_A(\widehat{\mathcal{U}}_A^\star,\widehat{\mathcal{U}}_B^\star) &\geq& 
\$_A(\widehat{\mathcal{U}}_A,\widehat{\mathcal{U}}_B^\star) \qquad   
\forall \quad
\widehat{\mathcal{U}}_A \\
\$_B(\widehat{\mathcal{U}}_A^\star,\widehat{\mathcal{U}}_B^\star) &\geq& 
\$_B(\widehat{\mathcal{U}}_A^\star,\widehat{\mathcal{U}}_B) \qquad   
\forall \quad
\widehat{\mathcal{U}}_B \quad . \nonumber\\
\nonumber
\end{eqnarray}

We define the novel function $\mathcal{N}_A$ of player A in a two-player quantum game by
\begin{eqnarray}
&&\mathcal{N}_A(\gamma):= \label{eq:qnf}\\
&&\int_{\theta_A=0}^{\pi}\int_{\varphi_A=0}^{\frac{\pi}{2}} 
\mathcal{N}_A(\widehat{\mathcal{Q}}_A^\star,\widehat{\mathcal{Q}}_B^\star,\theta_A,\varphi_A,\gamma) \, d\theta_A\, d\varphi_A -\nonumber\\
&&\int_{\theta_A=0}^{\pi}\int_{\varphi_A=0}^{\frac{\pi}{2}} 
\mathcal{N}_A(\widehat{\mathcal{D}}_A^\star,\widehat{\mathcal{D}}_B^\star,\theta_A,\varphi_A,\gamma) \, d\theta_A\, d\varphi_A  \,\, ,\nonumber
\end{eqnarray}
where the functions $\mathcal{N}_A(\widehat{\mathcal{Q}}_A^\star,\widehat{\mathcal{Q}}_B^\star,\theta_A,\varphi_A,\gamma)$ and $\mathcal{N}_A(\widehat{\mathcal{D}}_A^\star,\widehat{\mathcal{D}}_B^\star,\theta_A,\varphi_A,\gamma)$ are given by
\begin{eqnarray}
&\mathcal{N}_A(\widehat{\mathcal{U}}_A^\star,\widehat{\mathcal{U}}_B^\star,\theta_A,\varphi_A,\gamma)=&\\
&\mathcal{N}_A(\widehat{\mathcal{U}}_A^\star,\widehat{\mathcal{U}}_B^\star,\widehat{\mathcal{U}}_A,\gamma):=\$_A(\widehat{\mathcal{U}}_A^\star,\widehat{\mathcal{U}}_B^\star,\gamma) - 
\$_A(\widehat{\mathcal{U}}_A,\widehat{\mathcal{U}}_B^\star,\gamma) \quad .\nonumber&
\end{eqnarray} 
A rather lengthy calculation gives the following result for the function $\mathcal{N}(\gamma):=\mathcal{N}_A(\gamma)=\mathcal{N}_B(\gamma)$ of a two player quantum game with a prisoner's dilemma payoff matrix: 
\begin{eqnarray}
\mathcal{N}(\gamma)&=&\frac{\pi^2}{16} \left[ \left(1 + 3 \cos(2 \gamma) \right) 
\left( a - b \right) \right. \; +\; \nonumber\\
&&\left. \left( 5 - \cos(2 \gamma) \right) \left( c - d \right) \right] \quad .
\end{eqnarray}
An integration of $\mathcal{N}(\gamma)$ from $\gamma=0$ to $\gamma=\frac{\pi}{2}$ leads to a function ($\mathcal{N}$), that depends solely on the payoff parameters (a, b, c, d):
\begin{equation}
\mathcal{N}:=\int_{\gamma=0}^{\frac{\pi}{2}}\mathcal{N}(\gamma) \, d\gamma=\frac{\pi^3}{32} \left[ a-b+5\left( c-d \right) \right] \quad .
\end{equation}
In the following $\mathcal{N}$ will be used as the main cooperation indicator. 
It is easy to show, that the null of $\mathcal{N}(\gamma)$ is given by a specific threshold value of the entanglement: 
\begin{eqnarray}
\gamma_\star&:=&\left\{ \gamma \in [0,\frac{\pi}{2}]: \, \mathcal{N}(\gamma)=0 \right\} \nonumber\\
\gamma_\star&=&\frac{\pi}{2}\; -\; \frac{1}{2} \arccos \left(\frac{a\; -\; b\; +\; 5 \left( c\; -\; d \right)}{3 \left( a\; -\; b \right)\; -\; c\; +\; d} \right) \,\, . \label{eq:gamma_star}
\end{eqnarray}

In addition to $\mathcal{N}$ and $\gamma_\star$, two other other cooperation indicators are added: $\gamma_1$ is defined as the entanglement barrier, for which the classical Nash equilibrium $\left| DD \right>$ dissolves, and $\gamma_2$ is defined as the barrier where the new quantum Nash equilibrium $\left| QQ \right>$ appears.

\section{Classical vs. Quantum Cooperation Indicators}
The evolution of cooperation in repeated games depends on the payoff parameters of the game and the continuation probability $\delta$.\footnote{The present paper solely compares the classical theory of infinitely repeated games with the quantum approach. A comparison with the more general formulations based on negotiation and the axiomatic approach of two-player cooperative games \cite{nash_1953} will be addressed in a separate article.} Even though the theory of infinitely repeated games has been used to explain cooperation in a variety of environments it does not provide sharp predictions since there may be a multiplicity of equilibria \cite{dal_2006}.

In the classical theory of infinitely repeated games the standard lower bound on discount factors ($\underline{\delta}$) below which no player can ever cooperate on an equilibrium path of $\Gamma(\delta)$ depends simply on the payoff parameters $b, c$ and $d$: 
\begin{equation}\label{standard_delta}
\underline{\delta}:=\frac{b-c}{b-d} \quad.
\end{equation}

Blonski et al. define a new bound on the discount factors ($\delta^\star$), which includes the ''sucker's payoff'' (a)
\begin{equation}\label{blonski_delta}
\delta^\star:=\frac{b-a-c+d}{b-a} \quad,
\end{equation}
The authors of \cite{blonski_2007} show that this indicator is able to predict the cooperation percentage much better than the standard indicator $\underline{\delta}$.

\begin{figure}[h]
\centerline{
\includegraphics[width=3.5in]{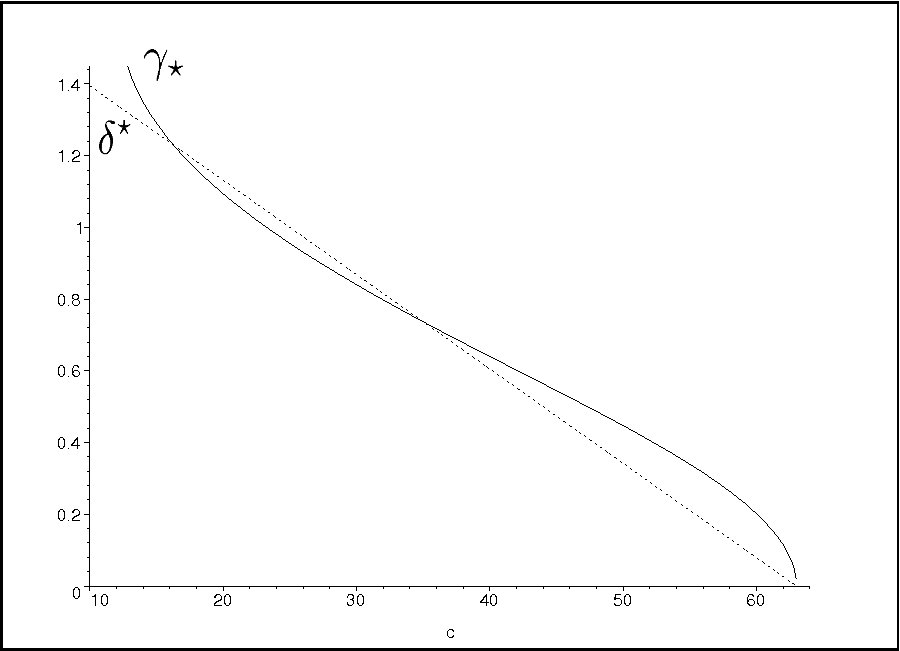}
}
\caption{$\delta^\star$ (dashed line, see \cite{blonski_2007}) and $\gamma_\star$ (solid curve) as a function of the payoff parameter c.}
\label{fig:gamma_delta}
\end{figure}

It is remarkable, that $\gamma_{\star}$ and $\delta^{\star}$ are for a wide range of possible payoff parameters quite similar. 
Figure \ref{fig:gamma_delta} illustrates the similarities of the functions $\gamma_{\star}$ (solid curve) and $\delta^{\star}$ (dashed line) by varying the parameter $c$ while keeping the other payoff parameters fixed as in the experimental settings of Dal B$\acute{\mbox{o}}$ et. al. \cite{dal_2006} ($a=12, b=50, d=25$).

\section{Experimental Validation}
\begin{figure}[h]
\centerline{
\includegraphics[width=3.5in]{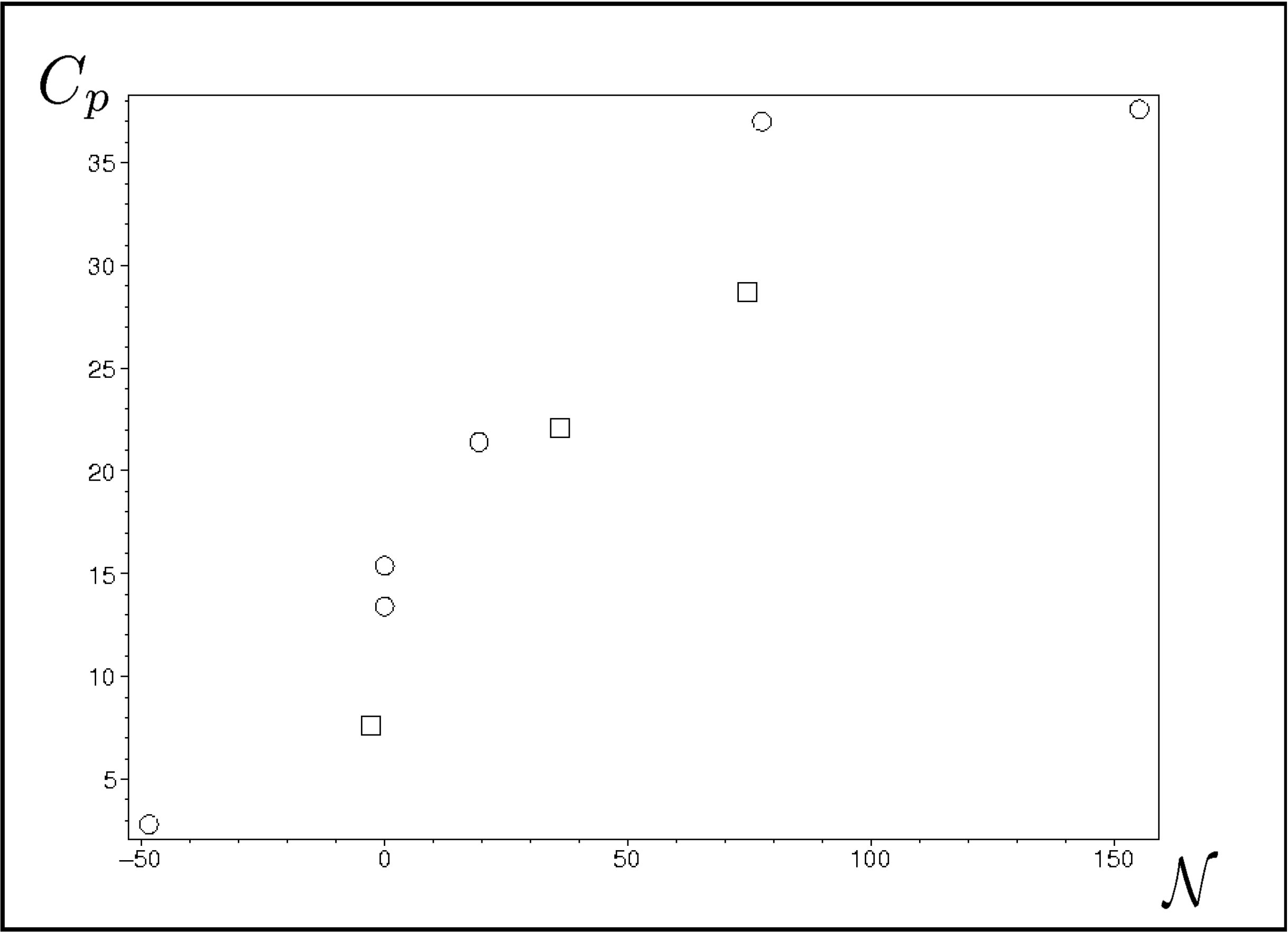}
}
\caption{Percentage of cooperating players ($C_p$) in experiment \cite{blonski_2007} (circles) and \cite{dal_2006} (boxes) as a function of $\mathcal{N}$.}
\label{fig:bexp_N}
\end{figure}

Quantum theoretical results of the games used in \cite{blonski_2007,dal_2006} and their experimental data is summarized in Table \ref{tab:results} and partly visualized in Figure \ref{fig:bexp_N}. 
The experimental data is based on the percentage of cooperating persons in all rounds. \footnote{For comparison reasons we have used the data with $\delta=0.75$ in both experiments. Besides the used payoff parameters, the discount factor $\delta$ is an important property in all of the experiments. As $\delta$ describes the abruption probability of the repeated games, it increases the individual entanglement of the persons, which results in a higher percentage of cooperation.}
In the sixth column of Table \ref{tab:results} the experimental findings of the percentage of cooperating persons ($C_p$) of Blonski et al. \cite{blonski_2007} and Dal B$\acute{\mbox{o}}$. et al. \cite{dal_2006} are denoted, whereas in the seventh column the cooperation rank of the games is quoted. 
The last rank in experiment \cite{blonski_2007} for example was found for game 2 ($C_p = 2.8 \%$), whereas the lowest cooperation rank was achieved in game 6 ($C_p = 37.6 \%$). 
\begin{longtable*}{|c|p{0.5cm}|p{0.5cm}|p{0.5cm}|p{0.5cm}|c|c|p{0.8cm}|p{0.8cm}||p{0.8cm}|p{0.8cm}|p{0.8cm}|p{1.0cm}|}
\caption[caption]{Quantum theoretical results and experimental data of Blonski et.al. \cite{blonski_2007} and Dal B$\acute{\mbox{o}}$ et al. \cite{dal_2006}.}
\label{tab:results}\\
\hline
\multicolumn{13}{|c|}{Experimental data of Blonski et. al. \cite{blonski_2007} and quantum theoretical results}\\\hline
Game No.& a  & b   & c  & d  & $C_p$   & Rank & $\underline{\delta}$ & $\delta^\star$ & $\gamma_2$ & $\gamma_1$ & $\gamma_\star$ & $\mathcal{N}$ \\\hline
1       & 70 & 100 & 90 & 80 & 21.4 \% & 3    & 0.5                  & 0.667          & 0.615      & 0.615      & 0.685          & 19.38         \\\hline
2       & 0  & 100 & 90 & 80 &  2.8 \% & 6    & 0.5                  & 0.9            & 0.322      & 1.107      & 0.866          & -48.45        \\\hline
3       & 30 & 130 & 90 & 70 & 15.4 \% & 4    & 0.667                & 0.8            & 0.685      & 0.685      & 0.785          & 0             \\\hline
4       & 0  & 100 & 90 & 70 & 13.4 \% & 5    & 0.333                & 0.8            & 0.322      & 0.991      & 0.785          & 0             \\\hline
5       & 0  & 120 & 90 & 50 & 37.0 \% & 2    & 0.429                & 0.667          & 0.524      & 0.702      & 0.685          & 77.52         \\\hline
6       & 0  & 140 & 90 & 30 & 37.6 \% & 1    & 0.625                & 0.786          & 0.641      & 0.481      & 0.615          & 155.03        \\\hline\hline
\multicolumn{13}{|c|}{Experimental data of Dal B$\acute{\mbox{o}}$ et. al. \cite{dal_2006} and quantum theoretical result}\\\hline
Game No.& a  & b   & c  & d  & $C_p$   & Rank & $\underline{\delta}$ & $\delta^\star$ & $\gamma_2$ & $\gamma_1$ & $\gamma_\star$ & $\mathcal{N}$ \\\hline
1       & 12 & 50  & 32 & 25 & 7.6 \%  & 3    & 0.72                 & 0.816          & 0.759      & 0.625      & 0.798          &  -2.91        \\\hline
2       & 12 & 50  & 40 & 25 & 22.1 \% & 2    & 0.4                  & 0.605          & 0.539      & 0.625      & 0.640          &  35.85        \\\hline
3       & 12 & 50  & 48 & 25 & 28.7 \% & 1    & 0.08                 & 0.395          & 0.231      & 0.625      & 0.487          &  74.61        \\\hline
\end{longtable*}
The next two subsequent columns in Table \ref{tab:results} present the lower bounds on the discount factors coming from standard ($\underline{\delta}$) and extended ($\delta^\star$) classical game theory. The last four columns sum up the specified cooperation indicators calculated with the use of quantum game theory. 
$\mathcal{N}$ is the most important indicator. Only if $\mathcal{N}$ is equal for two games the indicator $\gamma_\star$ should be used to classify the cooperation rank. In the games 3 and 4 of \cite{blonski_2007} neither $\mathcal{N}$ nor $\gamma_\star$ provide distinguisable values. In such a case one can use $\gamma_1$ and $\gamma_2$ to classify the cooperation rank, where $\gamma_1$ is expected to be more important than $\gamma_2$ because in real two-person games decisions depend firstly on the real strategy choices and only secondly on the imaginary part of the strategy choices. In game 3 the classical Nash equilibrium $\left| DD \right>$ disappears at $\gamma_1=0.685$, whereas in game 4 it vanishes at $\gamma_1=0.991$, which means that one expects to have more cooperating persons within game 3.

Figure \ref{fig:bexp_N} depicts the percentage of cooperating persons in both experiments as a function of $\mathcal{N}$. The diagram clearly shows, that an increase of cooperation comes along with an increase of $\mathcal{N}$. 

It should be mentioned that the comparison of two different experiments is difficult, because besides the fixed payoff parameters and the abruption rate $\delta$ other experimental details could influence the persons' cooperation behavior. For instance  the distribution of the persons strategic entanglement may depend on ethnological characteristics or maybe influenced by the experimental design. The information communicated by the experimenter himself could subliminally or even consciously influence the entanglement distribution of the whole group. Fig. \ref{fig:bexp_N} indicates a small difference between the mean of the persons' entanglement in both experiments, because the cooperation percentage in \cite{blonski_2007} is always somewhat above experiment \cite{dal_2006}. 

An increase (decrease) of $\delta$ influences the distribution of the players' entanglement, which results in an increase (decrease) of $C_p$. The strong correlation between $\mathcal{N}$ and $C_p$ for the specific games remains. 

Our work does not contradict the results of \cite{chen-2006}, but we presume, that by implementing a specific quantum version of the prisoner's dilemma game, the experimenters have increased the strength of entanglement of the players' strategic decisions (and as a result the cooperation percentage $C_p$). 

\section{Conclusion}
This article shows that a quantum extension of classical game theory is able to describe the experimental findings of two person prisoner's dilemma games. A classification scheme was introduced to evaluate the cooperation hierarchy of two-person prisoner's dilemma games. Four cooperation indicators where defined to predict the cooperation behavior. This quantum game theoretical approach was compared with predictions based on classical game theory and successfully tested for two experimental settings.

\section*{Acknowledgments}
We want to thank Peter Ockenfels and Matthias Blonski for helpful discussions.
This research is supported by grants from the German National Science Foundation (DFG) (Project “Scientific Publishing and Alternative Pricing Mechanisms”, Grant No. GZ 554922). We gratefully acknowledge the financial support.

\end{document}